\documentclass[12pt,psamsfonts]{amsart}
\usepackage{amsmath}
\usepackage{amsthm}
\usepackage{amssymb}
\usepackage{amscd}
\usepackage{amsfonts}
\usepackage{amsbsy}
\usepackage{setspace}
\usepackage{graphicx}
\usepackage{color}
\usepackage{calc}
\usepackage{subfigure}
\usepackage{geometry}
\usepackage{indentfirst}
\usepackage{tikz}
\usetikzlibrary{arrows, patterns}

\newtheorem {theorem} {Theorem}

\begin{document}

\title[]
{Periodic orbits of the two fixed centers problem with a variational gravitational field}
 \author{Fabao~Gao$^{1,2}$ and Jaume~Llibre$^{2}$}
 \address{${}^1$School of Mathematical Science, Yangzhou University, Yangzhou 225002, China}
  \address{\textnormal{E-mail: gaofabao@sina.com (Fabao Gao, ORCID 0000-0003-2933-1017})}
 \address{${}^2$ Departament de Matem$\grave{\text{a}}$tiques, Universitat Aut$\grave{\text{o}}$noma de Barcelona, Bellaterra 08193, Barcelona, Catalonia, Spain}
  \address{\textnormal{E-mail: jllibre@mat.uab.cat (Jaume Llibre, ORCID 0000-0002-9511-5999})}
  

\begin{abstract}
We prove the existence of periodic orbits of the two fixed centers problem bifurcating from the Kepler problem. We provide the analytical expressions of these periodic orbits when the mass parameter of the system is sufficiently small.
\end{abstract}

\maketitle

\section{Introduction}

\par The non-integrability of the restricted three-body problem prevents to obtain the analytical expressions of its general solutions. The periodic orbits of this problem have extremely important applications in practical space missions. This fact has attracted a large number of mathematicians and astronomers to carry out research on the periodic behavior of the classical restricted three-body problem (see \cite{Musielak} and the references therein). The extensive research covered three categories: qualitative analysis (see  \cite{Gao}, \cite{Gomez}, \cite{Koon}, \cite{Musielak}, and so on), analytical calculation (see \cite{Farquhar}, \cite{Richardson1}, \cite{Richardson2}), and numerical simulation (see \cite{Chenciner}, \cite{Henon}, \cite{Li}, \cite{Moore}, \cite{Simo}, \cite{Suvakov}).

\par For the planar circular restricted three-body problem, Zotos \cite{Zotos} investigated the problem with two equivalent masses with strong gravitational field, which was controlled by the power $p$ of gravitational potential. He revealed the great influence of the power $p$ on the nature of orbits. For the planar rotating Kepler problem, Llibre and Pa\c{s}ca \cite{Llibre-2} proved that some of the symmetric periodic orbits can be continued to the case of the restricted three-body problem colliding on a plane by using a continuation method. The method was also applied by Llibre and Makhlouf \cite{Llibre-1} to provide sufficient conditions for periodic orbits of a fourth-order differential system. For large values of the eccentricity, Abouelmagd et al. \cite{Abouelmagd} found that the anisotropic Kepler problem with small anisotropy has two periodic orbits in every negative energy level bifurcating from elliptic orbits of the Kepler problem by using averaging theory. In addition, they also presented the approximate analytic expressions of the continued periodic orbits. Recently, for each eccentricity and a sufficiently small parameter, Llibre and Yuan \cite{Llibre-3} continued elliptic periodic orbits of the Kepler problem to a hydrogen atom problem and an anisotropic Manev problem, respectively.

\par Some relevant works on the analytical calculation of the periodic orbits of the circular restricted three-body problem have been done by Farquhar and Kamel \cite{Farquhar}. They proposed an approximation method for computing periodic orbits, which motivated a plenty of research for computing periodic orbits of that problem. Based on the Lindstedt-Poincar\'e-like method and a successive approximation method, Richardson \cite{Richardson1}, \cite{Richardson2} constructed the classical periodic halo-type orbits of the circular restricted problem several years later. He gave a third-order analytical solution that was later on used in the orbital design.
 
\par Mainly motivated by Zotos \cite{Zotos} and Llibre \cite{Llibre-1} we shall use the fact that when the masses of two primaries differ greatly, i.e. the mass parameter of the two body centers problem is very small, then the two fixed centers problem becomes close to Kepler problem. In this paper we study which Keplerian periodic orbits can be continued, using the averaging theory, to the fixed-center problem with a variational gravitational field. Moreover we provide an analytical estimation of these periodic orbits.

The two body problem when one of the primaries is very big and the other is small, is very close to the restricted three-body problem with one of the primaries very big and the other small; these two problems have many interesting applications in astronomy, and in particular in the Solar system, see for instance \cite{G1, G2, Sz}.
 
 \section{Equations of Motion}

\par The equations of motion of a particle in the two fixed centers problem with a variational gravitational field can be written as
 \begin{equation}\label{1}
q'_k=\dfrac{\partial \mathcal{H}}{\partial p_k},\ \ \ p'_k=-\dfrac{\partial \mathcal{H}}{\partial q_k},\ \ \ \mbox{for}\ k=1\ \mbox{or}\ 2,
\end{equation}
 where
 \begin{equation}\label{1}
 \mathcal{H}=\dfrac{p^2_1+p^2_2}{2}-\dfrac{1-\mu}{r_1}-\dfrac{\mu}{r^p_2},
 \end{equation}
and $r_1=\sqrt{(q_1+\mu)^2+q^2_2}$, $r_2=\sqrt{(q_1-1+\mu)^2+q^2_2}$. Here the prime denotes derivative with respect to the time $t$, and $\mu$ ($0<\mu\ll1$) denotes the mass parameter of the two masses, one of mass $1-\mu$ fixed at $(-\mu,\,0)$ and the other of mass $\mu$ fixed at $(1-\mu,\,0)$. The parameter $p$ denotes the power of the gravitational potential. When $p$ is negative, the interaction is variational for different distances. Zotos in \cite{Zotos} studied the restricted circular three-body problem when $p>1$. Here we shall study the two fixed centers problem with $p=-2$. 

\par The main result of this paper is the following one.

\begin{theorem}\label{t1}
For the mass parameter $\mu>0$ sufficiently small and for every value $\mathcal{H}_0\in(-1/2,\,0)$, the periodic solution of the Kepler problem with elliptic eccentricity $e=-2\mathcal{H}_0$ can be continued to the two fixed centers problem with variational gravitational field $p=-2$.
\end{theorem}

 \section{Proof of theorem}
\par Introducing the so-called McGehee coordinate system and denoting it with $(r,\,\theta,\,v,\,u)$, where $r$ and $\theta$ are the radius and the angle in polar coordinates, $v$ and $u$ are the scaled components of velocity in the radial and angular directions, respectively. More precisely
\begin{equation*}
\begin{array}{rl} \vspace{2mm}
(q_1,q_2)&=\,r(\cos\theta,\ \sin\theta),\\
 \vspace{2mm}
 r^{-1/2}v&=\,(p_1,\ p_2)\cdot(\cos\theta,\ \sin\theta),\\
 r^{-1/2}u&=\,(p_1,\ p_2)\cdot(-\sin\theta,\ \cos\theta).
\end{array}
\end{equation*}

Then the Hamiltonian of system (2) with $p=-2$ becomes

\begin{equation}\label{3}
\begin{array}{rl} \vspace{2mm}
 \mathcal{H}=&\dfrac{p^2_1+p^2_2}{2} -\dfrac{1}{\sqrt{q^2_1+q^2_2}}\\ 
&+\mu\left(-1+2q_1-q^2_1-q^2_2+\dfrac{q_1+q^2_1+q^2_2}{\left(q^2_1+q^2_2\right)^{3/2}} \right)+\mathcal{O}(\mu^2),
\end{array}
\end{equation}

\noindent and we have

\begin{equation}\label{4}
\begin{array}{rl} \vspace{2mm}
 r'=&r^{-1/2}v,\\
  \vspace{2mm}
 \theta'=&r^{-3/2}u,\\
  \vspace{2mm}
 v'=&r^{-3/2}\left(u^2+\dfrac{v^2}{2}-1\right)\\
  \vspace{2mm}
 &+\mu\,r^{-5/2}\left[r\left(1+2r^3\right)+2\left(1-r^3\right)\cos\theta\right]+\mathcal{O}(\mu^2),\\
  \vspace{2mm}
 u'=&r^{-3/2}\left(-\dfrac{1}{2}uv\right)+\mu\,r^{-5/2}\left(1+2r^3\right)\sin\theta+\mathcal{O}(\mu^2).
\end{array}
\end{equation}

\par Note that $r=0$ corresponds to the collision singularity of equations (4). Performing the change $\mbox{d}t/\mbox{d}\tau=r^{5/2}$ in the independent variable, equations (4) can be rewritten as
\begin{equation}\label{5}
\begin{array}{rl}
 \vspace{2mm}
\dfrac{\mbox{d}r}{\mbox{d}\tau}=&r^2v,\\
 \vspace{2mm}
\dfrac{\mbox{d}\theta}{\mbox{d}\tau}=&ru,\\
 \vspace{2mm}
\dfrac{\mbox{d}v}{\mbox{d}\tau}=&r\left(u^2+\dfrac{v^2}{2}-1\right)\\ 
 \vspace{2mm}
 &+\ \mu\left[r\left(1+2r^3\right)+2\left(1-r^3\right)\cos\theta\right]+\mathcal{O}(\mu^2),\\
\dfrac{\mbox{d}u}{\mbox{d}\tau}=&r\left(-\dfrac{1}{2}uv\right)+\mu\left(1+2r^3\right)\sin\theta+\mathcal{O}(\mu^2).
\end{array}
\end{equation}

\par Consider that when $\mu=0$, the Hamiltonian system (3) is reduced to the Kepler problem. We shall study which periodic orbits of the Kepler problem can be continued to the two fixed centers problem with variational gravitational field $p=-2$ at a given energy level.

\par At a given energy level $\mathcal{H}=\mathcal{H}_0$ we have
\begin{equation}\label{6}
\dfrac{u^2+v^2}{2}-1+\mu\left[1-r-r^3+\dfrac{(1+2r^3)}{r}\cos\theta\right]=r\mathcal{H}_0.
\end{equation}

\par Isolating the radius $r$ from equation (6) and performing a series expansion with respect to the small mass parameter $\mu$ we get
\begin{equation}\label{7}
r=r(\theta,\ v,\ u,\ \mathcal{H}_0)=R_0+\mu R_1+\mathcal{O}\left(\mu^2\right),
\end{equation}
where
\begin{equation*}
\begin{array}{rl}\vspace{2mm}
 R_0=&\dfrac{1}{2\mathcal{H}_0}\left(u^2+v^2-2\right),\\\vspace{2mm}
 R_1=&\dfrac{1}{8\mathcal{H}_0^4}\left[8\mathcal{H}_0^3-4\mathcal{H}_0^2\left(u^2+v^2-2\right)-\left(u^2+v^2-2\right)^3\right]\\\vspace{2mm}
 &+\dfrac{1}{2\mathcal{H}_0^3}\dfrac{4\mathcal{H}_0^3+\left(u^2+v^2-2\right)^3}{u^2+v^2-2}.
\end{array}
\end{equation*}
Therefore substituting equation (7) into equations (5) and changing the independent variable $\tau$ by the variable $\theta$, equations (5) become
\begin{equation}\label{9}
\begin{array}{rl}\vspace{2mm}
\dfrac{\mbox{d}v}{\mbox{d}\theta}= &\dfrac{2u^2+v^2-2}{2u}+\mu\left[\dfrac{1}{u}+\dfrac{\left(u^2+v^2-2\right)^3}{4\mathcal{H}_0^3u}\right.\\\vspace{5mm}
&\hspace{24mm}+\left.\dfrac{8\mathcal{H}_0^3-\left(u^2+v^2-2\right)^3}{2\mathcal{H}_0^2\left(u^2+v^2-2\right)u}\cos\theta\right]+\mathcal{O}\left(\mu^2\right),\\\vspace{2mm}
\dfrac{\mbox{d}u}{\mbox{d}\theta}= &-\dfrac{v}{2}+\mu\dfrac{4\mathcal{H}_0^3+\left(u^2+v^2-2\right)^3}{2\mathcal{H}_0^2\left(u^2+v^2-2\right)u}\sin\theta+\mathcal{O}\left(\mu^2\right).
\end{array}
\end{equation}

\par If $\mu=0$ equations (8) will be reduced to the following unperturbed system
\begin{equation}\label{10}
\begin{array}{rl}\vspace{5mm}
 &\dfrac{\mbox{d}v}{\mbox{d}\theta}=\dfrac{2u^2+v^2-2}{2u},\\ 
 &\dfrac{\mbox{d}u}{\mbox{d}\theta}=-\dfrac{v}{2},
\end{array}
\end{equation}
which admits the general solution
\begin{equation*}
\begin{array}{rl}\vspace{5mm}
 &v(\theta;\,e,\,\theta_0)=\dfrac{e\sin(\theta-\theta_0)}{\sqrt{1+e\cos(\theta-\theta_0)}},\\
 &u(\theta;\,e,\,\theta_0)=\sqrt{1+e\cos(\theta-\theta_0)},
\end{array}
\end{equation*}
where $\theta_0\in[0,\ 2\pi)$ and $e$ are the argument of pericenter and the eccentricity of the Kepler problem, respectively. The value $e=0$ corresponds to circular periodic solutions, and $e\in(0,\ 1)$ indicates that the corresponding periodic solutions are elliptical ones.

\par In order to apply the averaging theory summarized in the Appendix to equations (8), we do the following transformations
\begin{equation*}
\begin{array}{ll}\vspace{5mm}
\mathbf{x}=\binom{v}{u},\ 
\mathbf{x}(\theta;\mathbf{z},0)=\binom{v(\theta;\ e,\ \theta_0)}{u(\theta;\ e,\ \theta_0)},\ 
\mathbf{z}=\binom{e}{\theta_0},\\
\mathbf{F_0}=\binom{F_{01}}{F_{02}},\ 
\mathbf{F_1}=\binom{F_{11}}{F_{12}},
\end{array}
\end{equation*}
where
\begin{equation*}
\begin{array}{rl}\vspace{2mm}
 F_{01}&=\dfrac{2u^2+v^2-2}{2u},\\
 \vspace{2mm}
 F_{02}&=-\dfrac{v}{2},\\
 \vspace{2mm}
 F_{11}&=\dfrac{1}{u}+\dfrac{\left(u^2+v^2-2\right)^3}{4\mathcal{H}_0^3u}+\dfrac{8\mathcal{H}_0^3-\left(u^2+v^2-2\right)^3}{2\mathcal{H}_0^2\left(u^2+v^2-2\right)u}\cos\theta,\\
 F_{12}&=\dfrac{4\mathcal{H}_0^3+\left(u^2+v^2-2\right)^3}{2\mathcal{H}_0^2\left(u^2+v^2-2\right)u}\sin\theta.
\end{array}
\end{equation*}

\par Now we calculate the averaged function $\mathcal{F}(\mathbf{z})$, see equation (15) in the Appendix.
\par Let 
\begin{equation}\label{10}
(G_1(\theta;\,e,\,\theta_0),\ G_2(\theta;\,e,\,\theta_0))=M_\mathbf{z}^{-1}(\theta,\,\mathbf{z})\ \mathbf{F_1}(\theta,\,\mathbf{x}(\theta;\,\mathbf{z},\ 0)),
\end{equation}
where 
\begin{equation*}
M_\mathbf{z}^{-1}(\theta,\,\mathbf{z})=
\begin{pmatrix}\vspace{3mm}
    y_1(\theta;\,e,\,\theta_0) &  y_2(\theta;\,e,\,\theta_0) \\ 
    y_3(\theta;\,e,\,\theta_0) &  y_4(\theta;\,e,\,\theta_0) \\  
  \end{pmatrix},
\end{equation*}
\noindent denotes the inverse of the fundamental matrix of equations (9) with
\begin{equation*}
\begin{split}
 y_1(\theta;\,e,\,\theta_0)=&\dfrac{(1+e\cos(\theta-\theta_0))^{1/2}}{2(1+e\cos\theta_0)^{3/2}}
 \left[2\cos\theta(1+e\cos\theta_0)\right.\\
 &+\left.e\sin\theta\sin\theta_0\right],\\
 y_2(\theta;\,e,\,\theta_0)=&-\dfrac{1}{8(1+e\cos(\theta-\theta_0))^{1/2}(1+e\cos\theta_0)^{3/2}}\\
 &\cdot\left[2\left(8+e^2\cos\theta+8e\cos\cos\theta_0\right)\sin\theta\right.\\
 & \ \ \ +2(4\cos\theta_0+3e\cos 2\theta_0)\sin 2\theta\\
 &\ \ \ +16e(2+\cos\theta)\sin^2(\theta/2)\sin\theta_0\\
 &\ \ \ +\left.6e^2\sin^2\theta\sin2\theta_0\right],\\
 y_3(\theta;\,e,\,\theta_0)=&\dfrac{(1+e\cos(\theta-\theta_0))^{1/2}}{2(1+e\cos\theta_0)^{1/2}}\sin\theta,\\
 y_4(\theta;\,e,\,\theta_0)=&\dfrac{4\cos\theta + e (\cos(2\theta-\theta_0)+3\cos\theta_0)}{4(1+e\cos(\theta-\theta_0))^{1/2}(1+e\cos\theta_0)^{1/2}}.
\end{split}
\end{equation*}

\noindent Hence, we obtain
\begin{equation}\label{11}
\begin{split}
 &G_1(\theta;\,e,\,\theta_0)\\
 &=\dfrac{1}{8(1+e\cos\theta_0)^{3/2}}\left[\left(4+\dfrac{\left(-1+e^2\right)^3}{\mathcal{H}_0^3(1+e\cos(\theta-\theta_0))^3}\right.\right.\\
 &\ \ \ +\dfrac{16\mathcal{H}_0\cos\theta(1+e\cos(\theta-\theta_0))}{-1+e^2}\\
 &\ \ \ \left.\cdot\left(1-\dfrac{(-1+e^2)^3}{8\mathcal{H}_0^3(1+e\cos(\theta-\theta_0))^3}\right)\right)\\
 &\cdot(2\cos\theta(1+e\cos\theta_0)+e\sin\theta\sin\theta_0)\\
&+\dfrac{2\mathcal{H}_0\sin\theta}{-1+e^2}\left(1+\dfrac{(-1+e^2)^3}{4\mathcal{H}_0^3(1+e\cos(\theta-\theta_0))^3}\right)\\
&\cdot\left(-2\left(8+e^2\cos\theta+8e\cos\theta_0\right)\sin\theta\right.\\
&-e(4\cos\theta_0+3e\cos2\theta_0)\sin2\theta\\
&-\left.\left.16e(2+\cos\theta)\sin\left(\dfrac{\theta}{2}\right)^2\sin\theta_0-6e^2\sin^2\theta\sin2\theta_0\right)\right],\\
&G_2(\theta;\,e,\,\theta_0)\\
 &=\dfrac{\sin\theta}{8(1+e\cos\theta_0)^{1/2}}
 \left[\left(4+\dfrac{\left(-1+e^2\right)^3}{\mathcal{H}_0^3(1+e\cos(\theta-\theta_0))^3}\right.\right.\\
 &+\dfrac{16\mathcal{H}_0\cos\theta(1+e\cos(\theta-\theta_0))}{-1+e^2}\\
 &\ \ \ \left.\cdot\left(1-\dfrac{(-1+e^2)^3}{8\mathcal{H}_0^3(1+e\cos(\theta-\theta_0))^3}\right)\right)\\
 &+\dfrac{4\mathcal{H}_0(4\cos\theta+e(\cos(2\theta-\theta_0)+3\cos\theta_0))}{-1+e^2}\\
 &\ \ \ \left.\cdot\left(1+\dfrac{(-1+e^2)^3}{4\mathcal{H}_0^3(1+e\cos(\theta-\theta_0))^3}\right)\right].
\end{split}
\end{equation}

\par Substituting equations (11) into equation (15) of the Appendix and computing the corresponding integrals we obtain
\begin{equation}\label{12}
\mathcal{F}(\mathbf{z})=(f_1(e,\,\theta_0),\ f_2(e,\,\theta_0)),
\end{equation}
where
\begin{equation}\label{13}
\begin{array}{rl}\vspace{5mm}
 f_1(e,\,\theta_0)=&\dfrac{3\sqrt{1-e^2}}{32\mathcal{H}_0^3(1+e\cos\theta_0)^{3/2}}
 \left[3e^2-8\mathcal{H}_0\right.\\\vspace{5mm}
 &\left.-4e(-1+2\mathcal{H}_0)\cos\theta_0+e^2\cos2\theta_0\right],\\
f_2(e,\,\theta_0)=&\dfrac{3e\sqrt{1-e^2}\sin\theta_0}{16\mathcal{H}_0^3\sqrt{1+e\cos\theta_0}}.
\end{array}
\end{equation}

\par The function $\mathcal{F}(\mathbf{z})$ in equation (12) is the averaged function of system (8). According to the averaging theory (see the Appendix) we must compute the simple zeros of the system $f_1(e,\,\theta_0)=0,\ f_2(e,\,\theta_0)=0$. Here we must consider $e\neq0$, otherwise $f_2(e,\,\theta_0)\equiv0$, and then we cannot obtain any information based on the averaging theory. From the equation $f_2(e,\,\theta_0)=0$ we get
$$\theta_0^1=0,\ \theta_0^2=\pi.$$

\par Case I: If $\theta_0^1=0$, we have $e=1$ or $e=2\mathcal{H}_0$ from the first equation of (13). However, both values of $e$ should be discarded because $e\in(0,\,1)$ and $\mathcal{H}_0<0$.

\par Case II: If $\theta_0^2=\pi$, we have $e=1$ or $e=-2\mathcal{H}_0$ from the first equation of (13), and $e=1$ is excluded because this case does not generate periodic solutions. Thus $e=-2\mathcal{H}_0$ with the $\mathcal{H}_0\in(-1/2,\,0)$.

\par Note that the Jacobian
\begin{equation*}
J=\left|
\begin{array}{cc}   
    \dfrac{\partial g_1}{\partial e} &  \dfrac{\partial g_1}{\partial \theta_0} \\ 
     \dfrac{\partial g_2}{\partial e} &  \dfrac{\partial g_2}{\partial \theta_0} \\  
  \end{array}
  \right|_{(e,\,\theta_0)=(-2\mathcal{H}_0,\,\pi)}
  =\dfrac{9(2\mathcal{H}_0-1)}{64\mathcal{H}_0^5}\neq0.
\end{equation*}

\par Therefore if $\mathcal{H}_0\in(-1/2,\,0)$, the obtained periodic solution
\begin{equation*}
(v(\theta;\,-2\mathcal{H}_0,\,\pi),\ u(\theta;\,-2\mathcal{H}_0,\,\pi))=\left(\dfrac{2\mathcal{H}_0\sin\theta}{\sqrt{1+2\mathcal{H}_0\cos\theta}},\ \sqrt{1+2\mathcal{H}_0\cos\theta}\right),
\end{equation*}
can be continued to the periodic solution of the two fixed centers problem with variational gravitational field $p=-2$ when the mass parameter $\mu$ is sufficiently small. For the energy $\mathcal{H}_0$, defined in (6), we plotted in Figure 1 the continued periodic orbits at the energy levels $\mathcal{H}_0=-1/4,-1/8$, and $-1/16$, respectively. As it is shown in Figure 1 the size of the periodic orbits decreases when the energy increases.

\begin{figure}[h]
  \begin{minipage}{170mm}
\centering
\includegraphics[width=8cm]{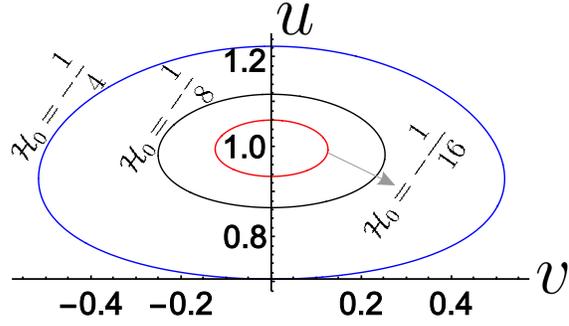}
   \caption{Periodic orbits at the energy level $\mathcal{H}_0=-1/4,-1/8$, and $-1/16$, respectively.}
  \end{minipage}
 \end{figure}
 
This completes the proof of Theorem \ref{t1}.

\section*{Appendix}
 
\par Consider the differential system 
\begin{equation}\label{14}
\dot{\mathbf{y}}=\mathbf{K_0}(t,\,\mathbf{y})+\mu\,\mathbf{K_1}(t,\,\mathbf{y})+\mu^2\,\mathbf{K_2}(t,\,\mathbf{y},\,\mu),
\end{equation}
with $\mu=0$ to $\mu\neq0$ small enough. The $\mathcal{C}^2$ functions $\mathbf{K_0},\ \mathbf{K_1}:\ \mathbb{R}\times\Omega\to\mathbb{R}^n$, $\mathbf{K_2}:\ \mathbb{R}\times\Omega\times(-\mu_0,\,\mu_0)\to\mathbb{R}^n$ are $T$-periodic with respect to variable $t$, and $\Omega$ is defined as an open subset of $\mathbb{R}^n$.

\par We assume that there is a submanifold $W$ of periodic solutions with the same period of the unperturbed system $\dot{\mathbf{y}}=\mathbf{K_0}(t,\,\mathbf{y})$. Let $\mathbf{y}(t,\,\mathbf{x},\,\mu)$ be the solution of this system such that $\mathbf{y}(0,\,\mathbf{x},\,\mu)=\mathbf{x}$. Then the linearized equation of the original unperturbed system along a periodic solution $\mathbf{y}(t,\,\mathbf{x},\,0)$ can be written as $\dot{\mathbf{z}}=D_{\mathbf{y}}\mathbf{K_0}(t,\,\mathbf{y}(t,\,\mathbf{x},\,0))\mathbf{z}$, and the fundamental matrix can be denoted by $M_\mathbf{x}(t)$. 

\par Suppose that there is an open set $W$ with $Cl(W)\subset\Omega$ such that $\mathbf{y}(t,\,\mathbf{x},\,0)$ is $\omega$-periodic for each $\mathbf{x}\in Cl(W)$ and $\mathbf{y}(0,\,\mathbf{x},\,0)=\mathbf{x}$. Here $Cl(W)$ denotes the closure of $W$ in $\mathbb{R}^n$. Then we have the following proposition (see Corollary 1 of \cite{Buica} for an easy proof):
\par \textbf{Proposition 1}. \textit{Let $W$ be an open and bounded set such that $Cl(W)\subset\Omega$ and for each $\mathbf{x}\in Cl(W)$ the solution $\mathbf{y}(t,\,\mathbf{x},\,0)$ satisfying $\mathbf{y}(0,\,\mathbf{x},\,0)=\mathbf{x}$ is $T$-periodic. Define the function $\mathcal{P}: Cl(W)\to\mathbb{R}^n$ as
\begin{equation}\label{15}
\mathcal{P}(\mathbf{x})=
\dfrac{1}{\omega}\int_{0}^{\omega}M_\mathbf{x}^{-1}(\theta,\ \mathbf{x})\ \mathbf{K_1}(\theta,\ \mathbf{y}(\theta;\ \mathbf{x},\ 0))\ \textnormal{d}\theta.
\end{equation}
Assume that there exists $\mathbf{c}\in W$ such that $\mathcal{P}(\mathbf{c})=\mathbf{0}$ and that $\textnormal{det}((\textnormal{d}\mathcal{P}/\textnormal{d}\mathbf{y})(\mathbf{c}))\neq 0$. Then system (14) admits an $\omega$-periodic solution $\psi(t,\,\mu)$ such that $\psi(0,\,\mu)\to \mathbf{c}$ for $\mu\to0$.}

 \section*{Acknowledgments}
 
We thank the reviewers and the editor-in-chief for their comments which helped us to improve the presentation of this paper.
 
 \par The first author gratefully acknowledges the support of the National Natural Science Foundation of China (NSFC) through grant No.11672259, the China Scholarship Council through grant No.20190832 0086.
 \par The second author gratefully acknowledges the support of the Ministerio de Econom$\acute{\i}$a, Industria y Competitividad, Agencia Estatal de Investigaci\'on grants MTM2016-77278-P (FEDER), the Ag$\grave{\text{e}}$ncia de Gesti\'o d'Ajuts Universitaris i de Recerca grant 2017SGR1617, and the H2020 European Research Council grant MSCA-RISE-2017-777911.

\end{document}